# *Topological Chiral-Gain in a Berry Dipole Material*

**Filipa R. Prudêncio[1,2*], and Mário G. Silveirinha[1]**

[1]University of Lisbon – Instituto Superior Técnico and Instituto de Telecomunicações,

Avenida Rovisco Pais 1, 1049-001 Lisbon, Portugal

[2]Instituto Universitário de Lisboa (ISCTE-IUL), Avenida das Forças Armadas 376, 1600-

077 Lisbon, Portugal

## Abstract

Recent studies have shown that non-equilibrium optical systems under static electric fields offer a pathway to realize chiral gain, where the non-Hermitian response of a material is controlled by the spin angular momentum of the wave. In this work, we uncover the topological nature of chiral gain and demonstrate how a static electric bias induces topological bandgaps that support unidirectional edge states at the material boundaries. Curiously, in our system, these topological edge states consistently exhibit dissipative properties. We further show that, by operating outside the topological gap, the chiral gain can be leveraged to engineer boundary-confined lasing modes with orbital angular momentum, locked to the orientation of the applied electric field. Our results open new possibilities for loss-compensated photonic waveguides, enabling advanced functionalities such as unidirectional, lossless edge-wave propagation and the generation of structured light with intrinsic orbital angular momentum.

---

[*] Corresponding author: filipa.prudencio@lx.it.pt

# I. Introduction

Topological metamaterials have revolutionized our understanding of material properties across diverse fields, ranging from condensed matter physics to photonics [1-9], by introducing robust and resilient states that are immune to local perturbations and to the influence of defects [10-13]. In nonreciprocal photonic platforms, topological phases are characterized by an integer number (the Chern number), which is a topological invariant insensitive to weak perturbations of the system's Hamiltonian, ensuring robustness in wave propagation and protecting against disorder-induced scattering. The topological classification of materials was initially developed for Hermitian systems. More recently, it was shown that non-Hermitian systems [14-26] —such as those exhibiting material absorption or material gain— can also display topological properties. Additionally, studies of the dispersive nature of nonreciprocal photonic materials have uncovered situations where traditional topological methods break down and the Chern topology becomes ill-defined [27-28]. The Chern topological numbers are typically calculated using topological band theory. Alternatively, the gap Chern number can be directly obtained from an integral of the photonic Green's function over a contour in the complex-frequency plane that contains the relevant band gap, avoiding the need for calculating the complete photonic band-structure [9, 26-30].

Nontrivial Chern insulators require a broken time-reversal symmetry, characteristic of nonreciprocal photonic platforms. The simplest way to break time reversibility is by incorporating into the system magneto-optic materials (e.g., ferrites and iron garnets) biased with a static magnetic field [31-32]. However, this approach requires bulky external biasing circuits. In recent years, several methods to achieve magnetless nonreciprocal responses have been studied [33-45]. However, there are still no truly competitive alternatives to the traditional magnetic bias solution, particularly in the infrared and optical domains.

In a recent work [46], it was demonstrated that the combination of material nonlinearities with a static electric bias results in highly nonreciprocal and non-Hermitian electromagnetic behaviors in low-symmetry materials. This theoretical investigation, inspired by semiconductor physics, suggests the feasibility of realizing bulk material responses qualitatively similar to those observed in transistor devices. Furthermore, a specific pathway to engineer these transistor-like responses was recently introduced, relying on a non-Hermitian electro-optic effect (NHEO) rooted in the Berry curvature dipole [47, 48]. The Berry curvature dipole is associated with a dipolar pattern of the electronic Berry curvature in the vicinity of the Fermi surface [49-51]. The electronic Berry curvature originates an "anomalous" term in the velocity of the Bloch electrons [49, 51], which leads to a nonlinear coupling between the external electric field and the quasi-momentum of the Bloch electrons. This nonlinear mechanism underpins the NHEO effect [52].

Importantly, the presence of a significant Berry curvature dipole necessitates the inversion symmetry breaking and is only possible in metallic systems where electronic bands are partially filled. This contrasts with insulating systems where an electric bias fails to induce a drift current crucial for unlocking non-Hermitian gain responses. Applying a static electric bias modifies the optical conductivity, breaking electromagnetic reciprocity and producing a non-Hermitian chiral-gain response [47, 48]. In fact, one of the most remarkable features of the NHEO effect is that the active or dissipative nature of the material response is dependent on the spin-angular momentum of the electromagnetic wave, which is related to the handedness of the polarization curve. For example, for waves polarized with a certain handedness the response may be active (exhibiting optical gain), while the opposite handedness results in a dissipative response.

Theoretical studies based on first principles functional density theory suggest that two-dimensional (2D) conductive materials, such as strained twisted bilayer graphene, or three-dimensional systems like trigonal tellurium, may serve as suitable candidates for achieving nonreciprocal and non-Hermitian transistor-like distributed responses with chiral-gain [47, 48]. Furthermore, theoretical and experimental studies have shown that applying an electric bias to natural trigonal tellurium, with the bias aligned along the trigonal axis, results in a gyrotropic nonreciprocal response [48, 53], which manifests in the so-called "kinetic" Faraday effect [54-55]. Consequently, trigonal tellurium, and more generally low-symmetry metallic-like materials with a large Berry curvature dipole hold significant potential for realizing electrically biased electromagnetic isolators and inducing pronounced optical dichroism [48].

Inspired by these recent developments, here we explore the topological properties of chiral gain media engineered through the NHEO effect. In particular, we calculate the topological phases of Berry dipole materials, such as natural trigonal tellurium, using Green's function methods. Our approach fully accounts for the material's intrinsic dispersion, dissipation mechanisms due to electron collisions, and the effects of an applied electric bias. Additionally, we characterize the dispersion of unidirectional edge-states propagating at interfaces between the chiral gain material and a trivial insulator. Finally, we demonstrate the potential of Berry dipole materials to engineer lasing modes with intrinsic orbital angular momentum, strictly locked to the orientation of the applied electric bias.

This article is organized as follows. In Section II, we briefly review the NHEO effect and the electro-optical response of generic low-symmetry conductive systems, with a focus on materials from the 32 point group, such as trigonal tellurium. We calculate the complex band structure of the bulk modes and examine the stability of the material as a function of the electric bias strength. In Section III, we investigate the topological phases of Berry dipole materials, demonstrating that a nontrivial bias opens a topological bandgap. Furthermore, we characterize the unidirectional edge-states that propagate at interfaces between the Berry dipole material and a trivial insulator. In Section IV, we characterize the lasing modes of a cavity formed by the Berry dipole material enclosed by metallic walls. We demonstrate that it is possible to engineer lasing modes confined to the cavity edges, with the propagation direction determined by the electric bias orientation. Finally, Section V summarizes the main findings of this work.

## II. NON-HERMITIAN LINEAR ELECTRO-OPTIC EFFECT

In this Section, we study electro-optic effects in low-symmetry metallic-type materials. As outlined in the Introduction, a static electric bias can modify the optical response of materials with a nontrivial Berry curvature dipole. This effect arises due to the so-called anomalous velocity of Bloch electrons which depends on the geometry of the electronic bands and on the electric field [49, 51]. As reviewed in Appendix A, the electron transport in low-symmetry metallic materials is governed by a (dimensionless) tensor $\bar{D}$, whose structure is dictated by the material's symmetry group.

For simplicity, here we focus on materials belonging to the 32 point group, e.g., trigonal tellurium [53]. Tellurium (Te) is a nonmagnetic chiral semiconductor that crystallizes in two mirror-image structures with space groups $P3_121$ and $P3_221$ [53]. The basic unit cell comprises three atoms arranged along a helical chain that spirals clockwise for the space group $P3_121$ and counterclockwise for the space group $P3_221$, with these chains forming a hexagonal lattice. The Berry curvature dipole tensor for Te and other materials belonging to the 32 point group is of the form [48, 53]:

$$\bar{\bar{D}} = \begin{pmatrix} D & 0 & 0 \\ 0 & D & 0 \\ 0 & 0 & -2D \end{pmatrix}. \tag{1}$$

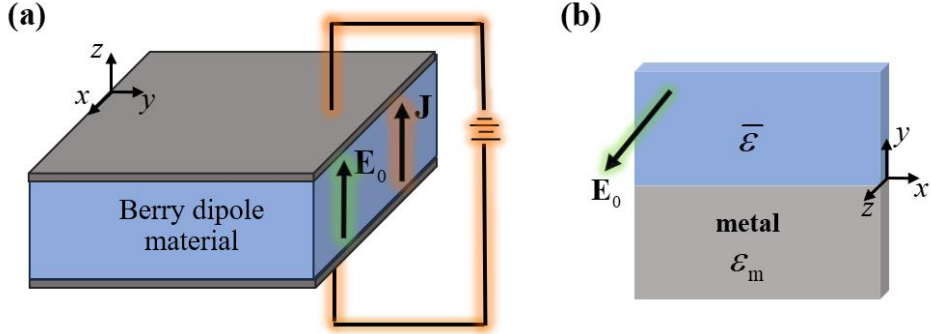

**Fig. 1 (a)** Berry dipole material biased with a static electric field $\mathbf{E}_0$ aligned with the $z$ direction. **(b)** Interface between the Berry dipole material ( $y > 0$ ) and a metal ( $y < 0$ ).

Throughout the article, we assume that the trigonal axis is aligned with the $z$-direction and that the static electric bias is applied along the trigonal axis, $\mathbf{E}_0 = E_0 \hat{\mathbf{z}}$ [see Fig. 1a]. In these conditions, the optical response of the material has the following gyrotropic-type structure [48]:

$$\bar{\bar{\varepsilon}} = \begin{pmatrix} \varepsilon_D & -i\varepsilon_{EO} & 0 \\ +i\varepsilon_{EO} & \varepsilon_D & 0 \\ 0 & 0 & \varepsilon_D \end{pmatrix}. \tag{2}$$

The diagonal terms ( $\varepsilon_D$ ) determine the response of the material without the electric-bias. For simplicity, we neglect anisotropic effects and natural optical activity, which may occur in low-symmetry materials. On the other hand, the non-diagonal terms ( $\varepsilon_{EO}$ ) represent the electro-optic response and arise due to the static bias. The two terms can be written explicitly as:

$$\varepsilon_\text{D} = 1 - \frac{1}{\omega}\frac{\omega_p^2}{\omega+i\Gamma}, \qquad \varepsilon_\text{EO} = \frac{\Gamma\omega_c}{\omega}\left(\frac{2}{\Gamma} + \frac{i\omega+\Gamma}{\Gamma^2+\omega^2}\right). \qquad (3)$$

In the above, $\omega_p$ is the plasma frequency and $\Gamma$ is the collision frequency. Furthermore, $\boldsymbol{\omega}_c = -\frac{e^3}{\varepsilon_0 \hbar^2 \omega_p^2}\left(\bar{D}^\text{T}\cdot \mathbf{E}_0\right) = 2\omega_c \hat{\mathbf{z}}$ is an equivalent oriented cyclotron frequency governed by the Berry dipole tensor. The scalar $\omega_c$ is given by $\omega_c = \frac{e^3}{\varepsilon_0 \hbar^2 \Gamma} DE_0$, with $\varepsilon_0$ the permittivity of vacuum, $c$ the speed of light and $\hbar$ the reduced Planck constant [48]. The magnetic response of the material is assumed trivial ($\bar{\mu} = \mu_0 \mathbf{1}_{3\times 3}$). It is worth to underline that the electrically induced gyrotropic optical response of tellurium was previously experimentally verified in Refs. [54-55]. For tellurium, both $\omega_p$ and $\Gamma$ lie in the terahertz range.

The permittivity component $\varepsilon_\text{EO}$ comprises two distinct contributions: one arising from conservative interactions (the first term in brackets in Eq. (3)), while the other stem from a non-Hermitian electrooptic response characterized by its non-conservative nature (second term in brackets) [48]. In tellurium, both terms contribute to the gyrotropic nonreciprocal response.

## A. Band structure and bulk stability

In this article, we focus on the propagation in the plane perpendicular to the trigonal axis of the material (*xoy* plane). The plane wave modes supported by the material are decoupled into transverse electric (TE) waves, with $E_z \neq 0$, and transverse magnetic (TM) waves, with $H_z \neq 0$. Evidently, only the TM waves ($\mathbf{E} = E_x\hat{\mathbf{x}} + E_y\hat{\mathbf{y}}$, $\mathbf{H} = H_z\hat{\mathbf{z}}$, and $\partial/\partial z = 0$) are sensitive to the electrically induced gyrotropy. Therefore, our analysis is centered on this case. The dispersion of these waves is governed by the characteristic equation [9]:

$$k^2 = \varepsilon_{\text{ef}} \left(\frac{\omega}{c}\right)^2, \qquad (4)$$

being $\mathbf{k} = k_x \hat{\mathbf{x}} + k_y \hat{\mathbf{y}}$ the wave vector, $\mathbf{k} \cdot \mathbf{k} = k^2 = k_x^2 + k_y^2$ and $\varepsilon_{\text{ef}} = \left(\varepsilon_D^2 - \varepsilon_{EO}^2\right)/\varepsilon_D$.

To evaluate the impact of the electric bias on the material's band structure and the stability of its optical response, we next examine the projected band structure for wave propagation in the *xoy* plane. The projected band structure represents the complex eigenfrequencies of the bulk modes for real-valued wave vectors. These eigenfrequencies are determined by solving Eq. (4) with respect to $\omega = \omega' + i\omega''$.

In the first example, in Fig. 2a we plot the projected band structure for the case where there is no bias ($\omega_c = 0$) corresponding to a material described by a dissipative Drude metal, with $\Gamma = 0.5\omega_p$. Due to the dissipation effects, the complex spectrum is completely confined in the lower-half frequency plane ($\omega'' < 0$). Note that we represent both the positive ($\omega' > 0$) and negative ($\omega' < 0$) parts of the frequency spectrum, which are linked by the particle-hole symmetry $\omega \to -\omega^*$. Additionally, a quasi-static branch of modes appears along the imaginary axis ($\omega' = 0$).

In Figs. 2b-2c, we illustrate the impact of the bias on the band structure of the material, for a weak and strong bias, respectively. The bias induces the opening of a band gap, as indicated by the two vertical pink-shaded strips. Specifically, the bias separates the original positive-frequency branch of the Drude model into two distinct components. In the limit of a weak bias, $\omega_c \to 0$ the two branches become the standard transverse and longitudinal bands of the unbiased plasma. The limiting points of each branch are marked in the dispersion curves with blue dots

–8–

($k = 0$) and black dots ($k \to \infty$). Note that for a trivial bias (Fig. 2a), some of the blue and black dots coincide because the longitudinal plasmons have a dispersion independent of $k$. The topological charge of the band gap will be characterized in the next section.

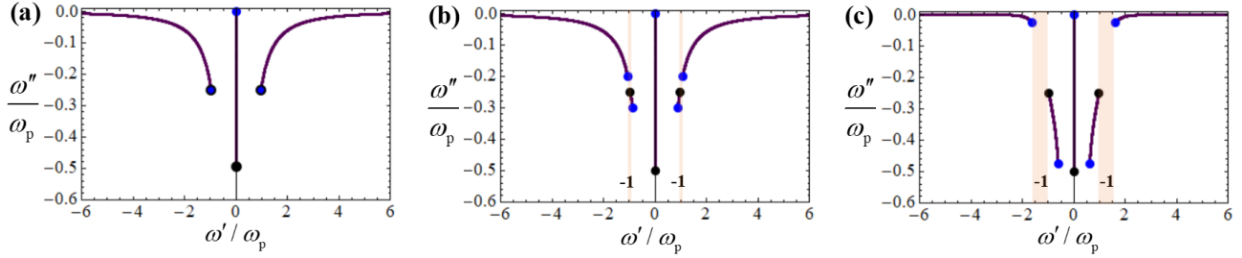

**Fig. 2** Projected band structure of the Berry-dipole material under the influence of an applied static electric field. The plots show a parametric representation of the complex-frequencies $\omega = \omega' + i\omega''$ as a function of the real-valued wave vector in the *xoy* plane. The collision frequency is $\Gamma = 0.5\omega_p$. The black dots represent the limiting points for $k \to \infty$, while the blue dots represent the limiting points for $k = 0$. **(a)** $\omega_c = 0$ (without the applied static electric field). **(b)** $\omega_c = 0.1\omega_p$ **(c)** $\omega_c = \omega_{c,th} \approx 0.5\omega_p$. The shaded vertical strip represents the relevant band gap, with gap Chern number -1.

Next, we investigate the conditions required to ensure a stable bulk response for wave propagation in the *xoy* plane. The material response is stable when the natural modes are in the lower-half of the complex frequency plane $\omega = \omega' + i\omega''$. We employ an approach similar to that used in Ref. [48], which analyzed the stability for propagation along the trigonal axis.

Specifically, we note that at the instability threshold, $\omega_{c,th}$, there must be an eigenmode with real valued frequency $\omega$ and real-valued wave vector $\mathbf{k}$. In other words, the band structure must "touch" the real-frequency axis. Thereby from Eq. (4), it follows that $\mathrm{Im}\{\varepsilon_{ef}(\omega)\} = 0$. Solving this equation with respect to $\omega_c$ we find that:

$$\left|\omega_{\mathrm{c}}(\omega)\right|=\sqrt{-\frac{\omega_{\mathrm{p}}^{6}+\left(\Gamma^{2}-2\omega_{\mathrm{p}}^{2}\right)\omega_{\mathrm{p}}^{2}\omega^{2}+\omega_{\mathrm{p}}^{2}\omega^{4}}{9\Gamma^{2}\omega_{\mathrm{p}}^{2}+\left(8\omega_{\mathrm{p}}^{2}-6\Gamma^{2}\right)\omega^{2}-4\omega^{4}}}\,. \tag{5}$$

The above formula gives the required bias strength so that the dispersion diagram intersects the real-frequency axis at frequency $\omega$. The instability threshold $\omega_{\mathrm{c,th}}$, is found by minimizing Eq. (5) with respect to $\omega$. Differentiating Eq. (5) with respect to $\omega$ and setting the result to zero, one finds that:

$$\omega \equiv \omega_{\mathrm{las}} = \sqrt{\frac{9\omega_{\mathrm{p}}^{2}}{2}+\frac{2\omega_{\mathrm{p}}^{4}}{\Gamma^{2}}+\frac{1}{2\Gamma^{2}}\sqrt{18\Gamma^{6}\omega_{\mathrm{p}}^{2}+16\omega_{\mathrm{p}}^{8}+57\Gamma^{4}\omega_{\mathrm{p}}^{4}+56\Gamma^{2}\omega_{\mathrm{p}}^{6}}}\,. \tag{6}$$

The frequency $\omega_{\mathrm{las}}$ gives the lasing frequency of the bulk material at the instability threshold. Figure 3a represents the bias strength [Eq. (5)] required to have an intersection at the frequency $\omega$. The minimum of the plot occurs at the lasing frequency $\omega_{\mathrm{las}}$. The plot assumes $\Gamma = 0.5\omega_{\mathrm{p}}$. For relatively small $\Gamma/\omega_{\mathrm{p}}$, the lasing frequency is roughly $\omega_{\mathrm{las}} \approx \omega_{\mathrm{p}}\sqrt{\frac{9}{2}+\frac{4\omega_{\mathrm{p}}^{2}}{\Gamma^{2}}} > 2.12\omega_{\mathrm{p}}$. Thus, the optical gain always arises near a frequency where the material behaves as a dielectric ($\mathrm{Re}\{\varepsilon_{\mathrm{D}}\} > 0$).

Evidently, the minimum of $\omega_{\mathrm{c}}$ occurs for $\omega_{\mathrm{c,th}} = \left|\omega_{\mathrm{c}}(\omega_{\mathrm{las}})\right|$. Substituting Eq. (6) into Eq. (5), we find that:

$$\omega_{\mathrm{c,th}} = \omega_{\mathrm{p}}\sqrt{\frac{4\omega_{\mathrm{p}}^{4}-\Gamma^{2}\omega_{\mathrm{p}}^{2}+3\Gamma^{4}+\left(4\omega_{\mathrm{p}}^{3}+3\Gamma^{2}\omega_{\mathrm{p}}\right)\sqrt{\omega_{\mathrm{p}}^{2}+2\Gamma^{2}}}{32\omega_{\mathrm{p}}^{4}+24\Gamma^{2}\omega_{\mathrm{p}}^{2}+18\Gamma^{4}}}\,. \tag{7}$$

The material response is stable provided $|\omega_c| < \omega_{c,th}$. It is clear from the previous equation that for small $\Gamma/\omega_p$, the cyclotron frequency at the instability threshold is $\omega_{c,th} \simeq 0.5\omega_p$. As shown in Fig. 3b, this estimate remains quite accurate even for fairly large values of $\Gamma/\omega_p$.

Figure 3c shows the projected band diagram of the material, for the cases of a stable bulk response ($|\omega_c| = 0.5\omega_{c,th}$), at the instability threshold ($|\omega_c| = \omega_{c,th}$), and for an unstable response ($|\omega_c| = 1.5\omega_{c,th}$), for the normalized collision frequency $\Gamma = 0.5\omega_p$. As seen, the response is stable for $|\omega_c| \leq \omega_{c,th}$ and the dispersion diagram crosses the real frequency axis nearby the lasing frequency $\omega_{las}$ (orange points).

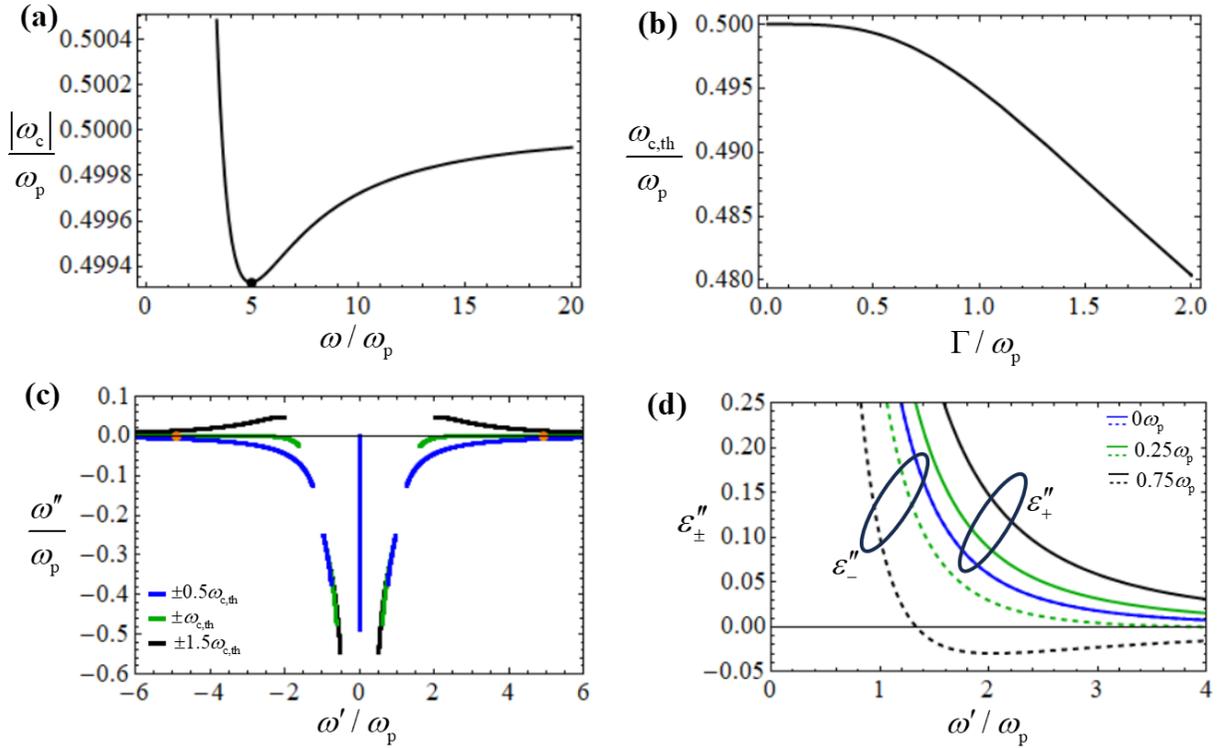

**Fig. 3** Study of the stability of the optical response. (a) Bias strength $|\omega_c|$ required to have a crossing at the real frequency $\omega$ [Eq. (5)]. The collision frequency is $\Gamma = 0.5\omega_p$. The black dot marks the lasing frequency $\omega_{las}$ for $\left|\omega_c\right| = \omega_{c,\text{th}}$ [Eq.(6)]. **(b)** Cyclotron frequency at the stability threshold as a function of the collision frequency $\Gamma$. **(c)** Projected band structure for a material with $\Gamma = 0.5\omega_p$ and (i) $\omega_c = \pm 0.5\omega_{c,\text{th}}$, (ii) $\omega_c = \pm\omega_{c,\text{th}}$, and (iii) $\omega_c = \pm 1.5\omega_{c,\text{th}}$. The orange points mark the lasing frequency $\omega_{\text{las}}$ at the instability threshold. **(d)** Eigenvalues of the non-Hermitian part of the material response $\bar{\varepsilon}'' = (\bar{\varepsilon} - \bar{\varepsilon}^\dagger)/(2i)$, for $\omega_c = 0$, $\omega_c = 0.25\omega_p$, and $\omega_c = 0.75\omega_p$.

## B. Non-Hermitian response

For non-conservative materials, it is useful to decompose the permittivity tensor as $\bar{\varepsilon} = \bar{\varepsilon}' + i\bar{\varepsilon}''$, where $\bar{\varepsilon}' = (\bar{\varepsilon} + \bar{\varepsilon}^\dagger)/2$ and $\bar{\varepsilon}'' = (\bar{\varepsilon} - \bar{\varepsilon}^\dagger)/(2i)$ are both Hermitian tensors. The tensor $\bar{\varepsilon}'$ governs the conservative part of the material response and determines the isofrequency contours of the bulk modes. In contrast, $\bar{\varepsilon}''$ describes the power dissipation in the material per unit of volume, given by $p_d = \frac{1}{2}\omega\varepsilon_0 \mathbf{E}^* \cdot \bar{\varepsilon}'' \cdot \mathbf{E}$. In conventional systems, the light-matter interactions are invariably dissipative, meaning that $p_d > 0$. In this case, the tensor $\bar{\varepsilon}''$ must be positive definite, with strictly positive eigenvalues. However, in systems with optical gain, the power $p_d$ can be negative because the material can supply energy to the wave.

A straightforward analysis shows that, for the optical response defined by Eqs.(2)-(3), the eigenvalues of the tensor $\bar{\varepsilon}''$ take the form:

$$\varepsilon''_+ = \frac{\Gamma \omega_p^2}{\Gamma^2 \omega + \omega^3} + \frac{\Gamma \omega \omega_c}{\Gamma^2 \omega + \omega^3}, \quad \varepsilon''_- = \frac{\Gamma \omega_p^2}{\Gamma^2 \omega + \omega^3} - \frac{\Gamma \omega \omega_c}{\Gamma^2 \omega + \omega^3}. \tag{8}$$

The corresponding eigenvectors are the two circular polarizations $\hat{\mathbf{e}}_+ = \frac{1}{\sqrt{2}}(\hat{\mathbf{x}} + i\hat{\mathbf{y}})$ and $\hat{\mathbf{e}}_- = \frac{1}{\sqrt{2}}(\hat{\mathbf{x}} - i\hat{\mathbf{y}})$. There is also an additional eigenvector $\hat{\mathbf{e}} \sim \hat{\mathbf{z}}$, corresponding to an electric field

–12–

oriented along the *z*-axis, but it is omitted here as we focus on TM waves, where the electric field is confined to the *xoy* plane.

As seen from Eq. (8), in the absence of the static electric bias ($\omega_c = 0$), the two eigenvalues are coincident and positive (blue line in Fig. 3d), typical of a passive response. As usual, the dissipation is rooted in the collisions of free-electrons with the ionic lattice.

Interestingly, under a nontrivial static bias ($\omega_c \neq 0$), the eigenvalues $\varepsilon_+''$, $\varepsilon_-''$ acquire an additional contribution. This corresponds to the second terms in Eq. (8), which arise from the linear electrooptic effect, and have *opposite* signs. Thereby, as the static bias strength increases, one of the circular polarizations experiences greater dissipation, while the opposite-handed circular polarization experiences reduced dissipation (see Fig. 3d, case $\omega_c = 0.25\omega_p$). Remarkably, for a strong enough bias one of the eigenvalues can become negative, corresponding to a situation where the material can experience gain. This is illustrated in Fig. 3d, for $\omega_c = 0.75\omega_p$. In this case, one of the eigenvalues ($\varepsilon_-''$) becomes negative (dashed black curve) while the other one ($\varepsilon_+''$) remains positive (solid black curve). This means that one of the circular polarizations ($\hat{\mathbf{e}}_+$) originates dissipation, while the circular polarization with opposite handedness ($\hat{\mathbf{e}}_-$) originates gain. This effect defines the material as a chiral gain medium. It is relevant to note that when the static bias is flipped, so that $\omega_c$ becomes negative, the eigenpolarizations that activate the gain and dissipation are interchanged.

It is useful to introduce the spin angular momentum of the electric field, defined as $\boldsymbol{\sigma} = i(\mathbf{E} \times \mathbf{E}^*)/|\mathbf{E}|^2$ [56]. The spin angular momentum is controlled by the handedness of the polarization curve. The spin angular momentum of the $\hat{\mathbf{e}}_\pm$ eigenfunctions is $\boldsymbol{\sigma} = \pm\hat{\mathbf{z}}$.

The part of the dissipated power $p_d$ associated with the non-Hermitian linear electrooptic effect can be written in terms of $\boldsymbol{\sigma}$ as:

$$p_{d,EO} = \frac{-1}{2}\omega\varepsilon_0 |\mathbf{E}|^2 \boldsymbol{\Omega}_\omega \cdot \boldsymbol{\sigma}, \quad \text{with} \quad \boldsymbol{\Omega}_\omega = -\frac{\omega_c \Gamma}{\Gamma^2 + \omega^2}\hat{\mathbf{z}}. \tag{9}$$

We refer to the dimensionless vector $\boldsymbol{\Omega}_\omega$ as the "gain vector". The formula above indicates that gain (or dissipative) interactions occur when the spin angular momentum of the wave is parallel (or anti-parallel) to the gain vector. This result generalizes the findings reported in Ref. [57] for dispersionless systems.

From Eq. (8), it is evident that for a chiral-chain configuration, where the eigenvalues $\varepsilon''_+, \varepsilon''_-$ have opposite signs, the bias strength must satisfy $|\omega_c| > \omega_p^2/\omega$ within a certain frequency range. This condition is fully compatible to the requirement for bulk stability ($|\omega_c| \leq \omega_{c,th}$), because for large frequencies $\omega_p^2/\omega$ can be arbitrarily small.

### III. TOPOLOGICAL PROPERTIES OF THE CHIRAL GAIN MEDIUM

In this section, we compute the topological invariants of the Berry dipole material with chiral gain, and characterize the corresponding topological edge states.

#### A. *Topological charge*

The spectrum of a system with a continuous translational symmetry is parameterized by a wave vector that "resides" on a plane. As a result, continuous systems without intrinsic periodicity generally lack a well-defined topological classification [9]. However, it has been shown that the response of such systems can always be regularized by introducing a high-spatial frequency cutoff, which suppresses nonreciprocal effects at short wavelengths ($k \to \infty$) [9].

Therefore, since the Berry dipole material is nonreciprocal, its regularized response can be associated with nontrivial topological phases.

The simplest way to account for material dispersion and non-Hermitian effects is to formulate the electromagnetic problem in a manner that explicitly incorporates the relevant physical degrees of freedom responsible for the dispersive response. In Appendix A, this is done for our system using a phenomenological transport equation that effectively describes how the material's free electrons are influenced by the external electric field and by the anomalous velocity term arising from the geometry of the electronic bands via the Berry dipole. This approach reduces the spectral problem to the canonical form $\mathbf{L_k} \cdot \mathbf{Q} = \frac{\omega}{c}\mathbf{Q}$, where $\mathbf{Q}$ is a state vector defined in Appendix A and $\mathbf{L_k}$ is the matrix operator in Eq. (A5).

The topological charge of a band gap can be expressed in terms of the system Green's function ($\mathcal{G}_\mathbf{k} = i\left(\mathbf{L_k} - \frac{\omega}{c}\mathbf{1}\right)^{-1}$) through an integral in the complex plane, taken over a line within the band gap that runs from $\omega_{\text{gap}} - i\infty$ to $\omega_{\text{gap}} + i\infty$ [26-30]. Here, $\omega_{\text{gap}}$ represents a frequency within the gap. The relevant gap in our system is shaded in pink in Figs. 2b and 2c, and takes the form $\omega_{\text{gap,L}} < \omega' < \omega_{\text{gap,H}}$ with $\omega_{\text{gap,L}} = \sqrt{\omega_\text{p}^2 - (\Gamma/2)^2}$. Another low-frequency gap is visible in the figure, but this gap is closed in the presence of a finite wave vector cutoff and, therefore, holds no topological significance (not shown).

Using the Green's function approach, we have determined that the high-frequency band gap is topological, characterized by the gap Chern number:

$$\mathcal{C}_{\text{gap}} = -\text{sgn}(\omega_\text{c}) = -\text{sgn}(E_0 D). \tag{10}$$

Thus, the sign of the topological charge is strictly tied to the sign of the product of the Berry dipole and to the orientation of the electric bias. By reversing the orientation of the electric bias, a topological phase transition can be triggered, causing the Chern number to reverse its sign.

## B. Topological edge-states

In Hermitian systems, the "bulk-edge correspondence" establishes a precise link between the gap Chern numbers of two topological materials and the net number of unidirectional edge states supported by a material interface [10-13]. In non-Hermitian systems, most notably in systems with gain, it has been shown that this correspondence may break down due to a phenomenon known as the "non-Hermitian skin effect" [19-22]. This effect occurs when the bulk spectrum undergoes a dramatic shift as the system transitions from periodic boundary conditions to "opaque" boundaries, commonly referred to as "open boundaries" in electronic systems. In systems exhibiting the non-Hermitian skin effect, bulk states tend to be exponentially localized at the boundary. To our knowledge, the non-Hermitian skin effect only occurs in systems with gain, with the bulk spectrum of the relevant operator spanning both the lower and upper halves of the frequency plane [23-24]. In particular, for systems with a spectrum confined to the lower-half frequency plane, the bulk-edge correspondence may remain valid.

To investigate the correlation between the gap Chern number and the emergence of topological edge states, we characterized the dispersion of boundary modes when the chiral-gain medium (region $y > 0$) is paired with a metal with permittivity $\varepsilon_\mathrm{m}$ (region $y < 0$), as shown in Fig 1b. The dispersion equation of the TM-polarized edge modes is described by [9]:

$$\frac{\varepsilon_\mathrm{D}^2 - \varepsilon_\mathrm{EO}^2}{\varepsilon_\mathrm{m}} \gamma_\mathrm{m} + \varepsilon_\mathrm{D} \gamma = \varepsilon_\mathrm{EO} k_x . \tag{11}$$

Here, $k_x$ is the propagation constant of the edge state, $\varepsilon_D, \varepsilon_{EO}$ are the diagonal and anti-diagonal elements of the permittivity tensor of the Berry dipole material [Eq. (3)], $\gamma = \sqrt{k_x^2 - \varepsilon_{ef}\left(\dfrac{\omega}{c}\right)^2}$ and $\gamma_m = \sqrt{k_x^2 - \varepsilon_m\left(\dfrac{\omega}{c}\right)^2}$ are the attenuation constants of the Berry dipole material and metal, respectively, along the y-axis, and $\varepsilon_{ef} = \left(\varepsilon_D^2 - \varepsilon_{EO}^2\right)/\varepsilon_D$.

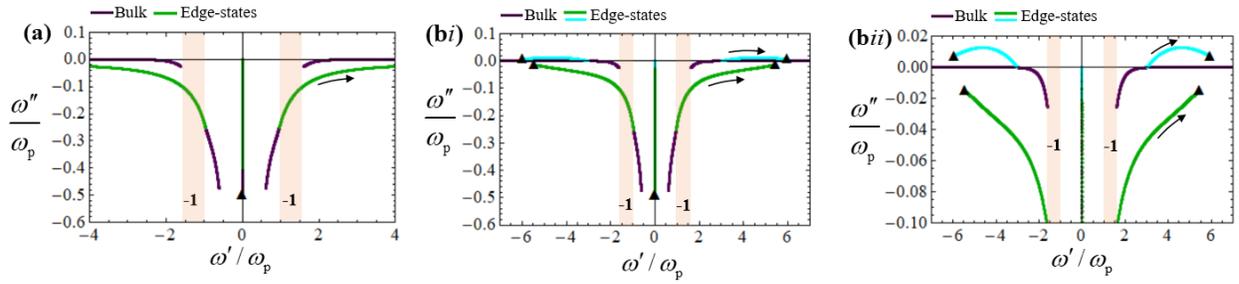

**Fig. 4** Complex dispersion of the edge-states (green and cyan lines) at the interface between the chiral-gain medium and a trivial medium (metal). The chiral gain medium is characterized by $\Gamma = 0.5\omega_p$. The black triangles mark the short-wavelength solutions ($k_x \to \pm\infty$) obtained with the quasi-static approximation. The black arrows indicate the direction of increasing $|k_x|$ along each branch of the edge-state dispersion curve. The purple lines represent the bulk modes. **(a)** Interface between a chiral gain medium with $\omega_c = 0.5\omega_p$ and a PEC. **(b$i$)** Interface between a chiral gain medium with $\omega_c = 0.5\omega_p$ and a low-loss metal with plasma frequency $\omega_m = 8\omega_p$. **(b$ii$)** Close-up of the dispersion of the edge-states depicted in panel bi). In all the panels, for $\omega' > 0$, the green lines are associated with edge states with a positive $k_x$, whereas the cyan lines are associated with edge states with a negative $k_x$.

For simplicity, in this section, we consider the metal as a perfect electric conductor (PEC), corresponding to $\varepsilon_m = -\infty$. In this case, the edge-state dispersion equation simplifies to $\varepsilon_D \gamma = \varepsilon_{EO} k_x$. Squaring both sides, some algebra reveals that the solutions of this equation must also satisfy $\varepsilon_D \left(\dfrac{\omega}{c}\right)^2 = k_x^2$. However, the reverse is not true; some solutions of the simplified equation do not satisfy the original dispersion. Interestingly, since $\varepsilon_D$ is independent of the bias in our problem, we find that the edge-state dispersion is also independent of $\omega_c$!

Figure 4a represents a parametric plot of the complex eigenfrequencies $\omega' + i\omega''$ of the edge-states as a function of $k_x$ real-valued. This dispersion models an edge waveguide that is closed on itself in the form of a loop. When $\omega_c > 0$, the solutions with $\text{Re}\{\omega\} > 0$ are linked to $k_x > 0$, indicating that the edge modes are unidirectional and propagate along the $+x$-axis. The edge state has a finite lifetime ($\text{Im}\{\omega\} < 0$) because of the material absorption. As shown in Fig. 4a, the edge states dispersion is gapless (green curve) and connects the two bulk bands, in accordance with the bulk-edge correspondence. Moreover, the propagation direction for the edge state (along $+x$) is also consistent with the sign of the gap Chern number ($\mathcal{C}_{\text{gap}} = -1$), as predicted by the bulk edge correspondence for Hermitian systems [13]. Note that, similar to the bulk case, there exists a branch of static-like edge-state solutions confined to the imaginary axis.

A few observations are in order. First, for simplicity, we neglect the effect of the high-frequency spatial cutoff in the calculation of the edge states. This could be accounted for using an approach similar to that described in Ref. [12], which involves additional boundary conditions at the PEC interface, specifically the vanishing of the electric current density **j** defined in Appendix A. The second observation pertains to the limit $\omega_c \to 0$. As mentioned, the edge-state dispersion is independent of bias strength but depends on the sign of $\omega_c$. For positive (negative) $\omega_c$, allowed edge states with $\text{Re}\{\omega\} > 0$ exhibit $k_x > 0$ ($k_x < 0$), implying a discontinuous transition at $\omega_c = 0$. This behavior arises because the linearized model of the material response assumes a static bias much stronger than the dynamical signal [48], making the model ill-defined in the limit $\omega_c \to 0$. A final observation is that, since the edge states for a PEC interface are insensitive to the bias strength, it follows that even if the bulk material is unstable, the dispersion of the topological edge state always lies in the lower-half frequency plane. Thus, for this configuration, it is impossible to exploit the gain properties of the Berry dipole material to induce amplifying unidirectional edge states.

## IV. BOUNDARY-CONFINED LASING MODES

In the following, we explore the potential of using the Berry dipole medium to realize boundary-confined lasing modes. The idea is to leverage the chiral-gain properties of the medium to excite amplifying edge states propagating along its boundary, while maintaining a stable bulk response (see Fig. 5a). As discussed in Sect. IIII.B, this cannot be achieved by enclosing the material with a PEC wall. Next, we demonstrate that a plasmonic boundary can achieve this, enabling the excitation of unidirectional edge states with gain.

### A. Gain momentum locking

It is well-known that surface plasmons in metals possess an intrinsic spin determined by their direction of propagation [56, 58, 59]. Specifically, the direction of the spin angular momentum $\boldsymbol{\sigma}$ of the surface plasmons is determined by the cross product of the momentum ($\mathbf{k}$) and the attenuation direction ($\boldsymbol{\gamma}$). This property, known as spin-momentum locking [56, 58, 59], implies that plasmons propagating in the counter-clockwise direction with respect to the *z*-axis have their spin angular momentum pointing along the positive *z* axis, while those propagating clockwise direction have their spin angular momentum oriented along the *–z* axis.

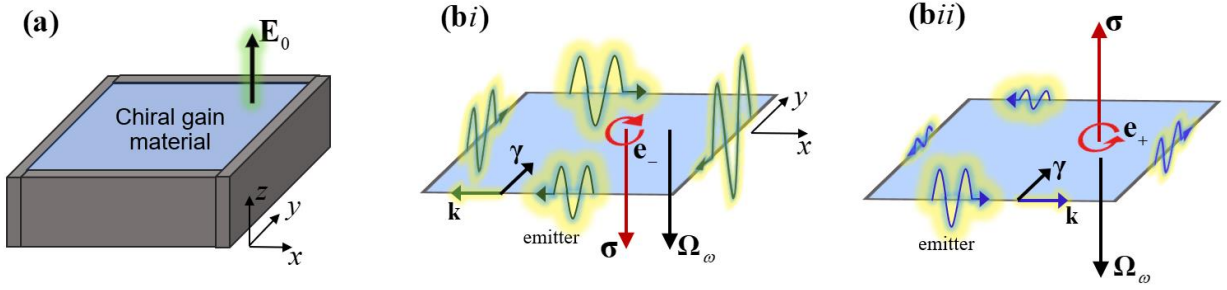

**Fig. 5 (a)** Chiral-gain medium cavity with metallic (plasmonic) walls. **(b)** Illustration of gain momentum locking. **(b*i*)** A surface plasmon propagating along the clockwise direction around the cavity walls has an intrinsic spin angular momentum **σ** directed along the –z axis. For $\omega_c > 0$, **σ** is parallel to the gain vector $\mathbf{\Omega}_\omega$ resulting in optical gain. **(b*ii*)** Similar to b*i*) but for a surface plasmon propagating in the counterclockwise direction. In this case, **σ** is anti-parallel to the gain vector $\mathbf{\Omega}_\omega$, resulting in increased dissipation.

Interestingly, as discussed in Sect. II.B, in a chiral gain medium, one spin state is associated with material dissipation [Fig. 5bi], whereas the opposite spin state leads to material gain [Fig. 5bii]. Thereby, as first proposed in Ref. [57], the interplay between spin-momentum locking and chiral gain results in gain-momentum locking: plasmons with a propagation direction such that the spin angular momentum is parallel to the gain vector $\mathbf{\Omega}_\omega$ are amplified within the medium [Fig. 5bi], while those with opposite propagation direction experience absorption [Fig. 5bii].

It is important to note that for $\omega_c > 0$, the gap Chern number is negative, meaning that if there is a single topological state, the bulk-edge correspondence dictates it must propagate in the counterclockwise direction within the cavity geometry [13]. However, from Eq. (9), we see that for $\omega_c > 0$, the gain vector is directed along the –z axis, i.e., anti-parallel to the plasmon spin angular momentum. Thus, in this system, the topological state is necessarily associated with dissipation. This offers a new perspective on the findings in Sect. III.B and further explains why the single topological state supported by the chiral-gain medium and a PEC wall is unaffected by optical gain.

## B. Unidirectional edge-states with optical gain

From the previous subsection, it is clear that for the considered dispersion model of the Berry dipole material, optical gain can only be observed in a spectral region outside the topological bandgap. Additionally, the cavity walls cannot be perfect conductors. Taking this into account, we now assume the cavity walls follow a dispersive Drude model described by the permittivity $\varepsilon_m(\omega) = 1 - \omega_m^2/\omega^2$ where $\omega_m$ is the plasma frequency of the metal. For simplicity, we assume that the dissipation in the cavity walls is negligible in this analysis.

Furthermore, similar to previous examples, we assume that the Berry dipole material is characterized by $\Gamma = 0.5\omega_p$, and we suppose that the bias strength is slightly below the instability threshold $\omega_{c,th} \approx 0.5\omega_p$ to maximize the gain effect. It should be noted that for the case of *p*-doped tellurium this bias strength is about 100 times larger than what has been considered in experiments [48]. This problem can be alleviated by using *n*-doped tellurium or other engineered material with a much stronger Berry dipole [48]. The metal is characterized by $\omega_m = 8\omega_p$.

The calculated edge-state dispersion [computed from Eq. (11)] is depicted in Fig. 4bi (green and cyan lines). It corresponds to a parametric plot of the complex eigenfrequencies $\omega' + i\omega''$ as a function of $k_x$ real-valued. The purple lines in Fig. 4bi represent the bulk dispersion, which is totally confined to the lower-half frequency plane. As seen, now there are two branches of edge states with $\omega' > 0$. The black triangles mark the short-wavelength solutions ($k_x \to \pm\infty$), which are obtained using a quasi-static formalism described in Appendix B.

Similar, to the example of Sect. III.B, one of the edge-state branches (colored in green) represents the gapless topological modes. In agreement with previous considerations, it is fully contained in the lower-half frequency plane. Remarkably, in the frequency window $3\omega_p \leq \omega' \leq 6\omega_p$, there is an additional branch (colored in cyan) that is partially contained in the upper-half frequency plane ($\omega'' > 0$) resulting in optical gain (see a zoom in Fig. 4bii). In agreement with the gain-momentum locking principle, this branch of edge states is associated with plasmons that propagate in the clockwise direction (–x axis for the interface y=0 in Fig. 1b). In the frequency window, $3\omega_p \leq \omega' \leq 6\omega_p$, one of the eigenvalues of $\bar{\varepsilon}''$ is negative [see Fig. 3d], confirming that the Berry dipole material has a chiral gain response.

Remarkably, the orbital angular momentum of the lasing mode is intrinsically linked to the orientation of the bias electric field. This makes the proposed system useful for generating structured light with inherent orbital angular momentum. Moreover, reversing the direction of the bias electric field $\mathbf{E}_0$, cause the lasing mode to propagate in the opposite direction. Lasing relying on bulk modes in Berry dipole materials was previously discussed in Refs. [60, 61].

To further illustrate these ideas, we calculated the fields associated with the edge states for the interface y=0 (Fig. 1b). The magnetic field complex amplitude is given by:

$$H_z(x,y) = H_0 e^{ik_x x} \begin{cases} e^{-\gamma y}, & y > 0 \\ e^{+\gamma_m y}, & y < 0 \end{cases}, \qquad (12)$$

Figure 6b represents a time snapshot of the field profile of the two edge states with $\omega/\omega_p = 4$, for the same material parameters as in Fig. 4b. Here, we assume open boundary conditions so that $\omega$ is taken as real-valued, whereas the propagation constant of the edge state $k_x$ is complex-valued. Consistent with the chiral gain response of the material and with the gain momentum locking, plasmons that propagate along the –x direction have a spin that matches the eigenpolarization that activates the gain resulting in amplification [Fig. 6bi]. On the other hand, plasmons that propagate along the +x direction activate dissipation in the material resulting in absorption [Fig. 6bii].

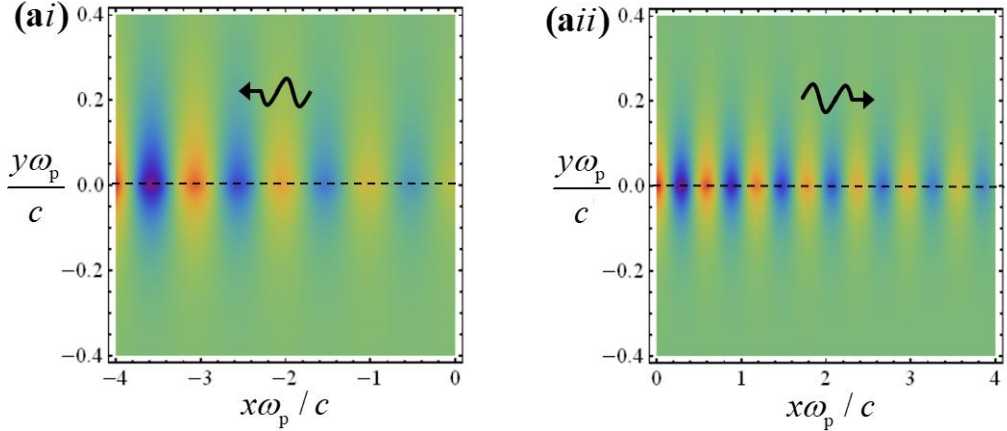

**Fig. 6** Density plot of the magnetic field associated with the boundary modes for the interface $y=0$. The black arrows indicate the direction of propagation of the edge state. (**a***i*) Plasmon propagating in the -*x* direction resulting in amplification. (**a***ii*) Plasmon propagating in the +*x* direction resulting in attenuation.

## V.  Conclusions

In this work, we uncovered the topological origin of the non-Hermitian linear electro-optic effect in systems with chiral-gain. For metallic-type materials from the 32 point group, such as trigonal tellurium, we demonstrated that applying an electric bias induces an equivalent cyclotron frequency, which opens a topological bandgap. Using a Green's function approach, we numerically calculated the gap Chern numbers, confirming the system's nontrivial topology.

We also investigated the dispersion of edge states at the interface between the Berry dipole material and a metal. Interestingly, our findings show that topological edge states in this system invariably exhibit dissipative properties—a characteristic that appears intrinsic to the model. However, by operating outside the topological gap, we were able to engineer unidirectional edge states with optical gain while maintaining stable bulk mode conditions. These edge states hold potential for designing lasing modes confined to material boundaries, producing structured light with intrinsic orbital angular momentum.

While in our system, a nontrivial topology is always associated with dissipative edge states, this raises an intriguing question: Is this property universal, or could other dispersive models, particularly those based on materials from different symmetry groups, generate topological edge states that take advantage of chiral gain without the same dissipation? This opens interesting avenues for future research.

In terms of practical implementation, the required bias strength to achieve boundary-confined lasing modes may be challenging for tellurium. However, other engineered materials with stronger Berry dipole effects could significantly reduce the necessary bias. Furthermore, for the excitation of edge modes, the bias only needs to be applied along the footprint of surface plasmons, which is essentially a line—making this approach more feasible in real-world applications. If successful, these systems could enable the development of loss-compensated waveguides for terahertz and far-infrared frequencies, potentially offering transformative advances in photonics.

### Acknowledgements

This work was supported in part by the IET under the A F Harvey Engineering Research Prize, by the Simons Foundation Award Number 733700, and by Fundação para a Ciência e a Tecnologia and Instituto de Telecomunicações under Project No. UIDB/50008/2020.

## Appendix A: Transport equation for Berry dipole materials

To characterize the topology of the dispersive optical material, it is useful to employ a Schrödinger-type dynamical evolution. This is mostly convenient done using the phenomenological model introduced in Ref. [52]. Specifically, it has been shown that the electrodynamics of a Berry dipole material can be effectively described by the following transport equation:

$$\frac{d\mathbf{p}}{dt} + \Gamma \mathbf{p} = q\mathbf{E}, \tag{A1a}$$

Here, $q = -e$ is the electron charge, $\Gamma$ is the collision frequency, $\mathbf{p}$ represents the (dynamical) "quasi-momentum" of the free-electrons and $\mathbf{E}$ is the (dynamical) electrical field. Unlike in conventional materials, where the electron velocity $\mathbf{v}$ is dictated by the curvature of the energy dispersion through the effective mass $m^*$, in low-symmetry crystals, an additional velocity contribution arises from the interaction between the Bloch electrons and the crystal lattice. This interaction, known as the anomalous velocity, is proportional to the electric field [49, 51]. The linearized effective velocity of the Bloch electrons can be described by [52]:

$$\mathbf{v} = \frac{\mathbf{p}}{m^*} + \frac{1}{m^*}\left(\bar{\bar{\zeta}} \cdot \mathbf{p}_0\right) \times \mathbf{E} + \frac{1}{m^*}\left(\bar{\bar{\zeta}} \cdot \mathbf{p}\right) \times \mathbf{E}_0. \tag{A1b}$$

The last two terms represent the linearized anomalous velocity contribution. Here, $\mathbf{E}_0$ is the static-electric bias, $\mathbf{p}_0 = q\mathbf{E}_0/\Gamma$ is the quasi-momentum for the static case, and the tensor $\bar{\bar{\zeta}}$ is proportional to the Berry curvature dipole, $\bar{\bar{\zeta}} = \frac{qm^*}{n\hbar^2}\bar{\bar{\mathbf{D}}}^T$, where $n$ is the free-electron density.

This transport equation is coupled to the Maxwell's equations in the usual manner:

$$\nabla \times \mathbf{E} = -\mu_0 \partial_t \mathbf{H}, \qquad \nabla \times \mathbf{H} = \mathbf{j} + \varepsilon_0 \partial_t \mathbf{E}, \tag{A1c}$$

where $\mathbf{j} = nq\mathbf{v}$ is the current density. The electrodynamics predicted by this model is fully consistent with Ref. [48].

Let us now consider a two-dimensional scenario ($\partial/\partial z = 0$), with $\mathbf{H} = H_z \hat{\mathbf{z}}$ and the remaining fields in the *xoy* plane ($\mathbf{E} = E_x \hat{\mathbf{x}} + E_y \hat{\mathbf{y}}$, etc). Furthermore, we suppose that the electric bias is along the $z$ direction ($\mathbf{E}_0 = E_0 \hat{\mathbf{z}}$) and that the Berry curvature dipole is as in Eq. (1). Then, a straightforward analysis shows that the electric current density can be expressed as:

$$j_x = \varepsilon_0 \omega_p^2 \tilde{p}_x - \omega_c \varepsilon_0 \Gamma \tilde{p}_y - 2\omega_c \varepsilon_0 E_y$$
$$j_y = \omega_c \varepsilon_0 \Gamma \tilde{p}_x + \varepsilon_0 \omega_p^2 \tilde{p}_y + 2\omega_c \varepsilon_0 E_x \quad , \quad (A2)$$

where $\omega_c = \dfrac{e^3}{\varepsilon_0 \hbar^2 \Gamma} DE_0$, $\omega_p^2 = \dfrac{e^2 n}{\varepsilon_0 m^*}$ and $\tilde{p}_x = p_x/q$ and $\tilde{p}_y = p_y/q$. Plugging the above formulas into the Maxwell's equations [Eq. (A1c)] and using the transport equation (A1a), one finds that the system dynamics reduces to:

$$\mathbf{L} \cdot \mathbf{Q} = \frac{1}{c} i \partial_t \mathbf{Q}. \quad (A3)$$

with $\mathbf{Q} = \begin{bmatrix} E_x & E_y & \eta_0 H_z & \tilde{p}_x & \tilde{p}_y \end{bmatrix}^T$ a state vector formed by the electromagnetic fields and the normalized quasi-momentum distribution, with $\eta_0$ the vacuum impedance. In the above, $\mathbf{L}$ represents the differential operator:

$$\mathbf{L} = i \begin{pmatrix} 0 & 2\omega_c & c\partial_y & -\omega_p^2 & \omega_c \Gamma \\ -2\omega_c & 0 & -c\partial_x & -\omega_c \Gamma & -\omega_p^2 \\ c\partial_y & -c\partial_x & 0 & 0 & 0 \\ 1 & 0 & 0 & -\Gamma & 0 \\ 0 & 1 & 0 & 0 & -\Gamma \end{pmatrix}. \quad (A4)$$

where $\partial_x$ and $\partial_y$ are the spatial derivatives.

The above theory can be easily modified to take into account the effect of a high-spatial frequency cutoff [9]. Specifically, this can be done by replacing the electric current density in the Maxwell's equations [Eq. (A1c)] as $\mathbf{j} \to \left(1 - \nabla^2 / k_{max}^2\right)^{-1} \mathbf{j}$ where $k_{max}$ determines the spatial cut-off. This leads to following operator in the spectral domain ($\nabla \to i\mathbf{k}$):

$$\mathbf{L}_\mathbf{k} = \begin{pmatrix} 0 & \dfrac{2i\omega_c}{1+k^2/k_{max}^2} & -ck_y & \dfrac{-i\omega_p^2}{1+k^2/k_{max}^2} & \dfrac{i\omega_c \Gamma}{1+k^2/k_{max}^2} \\ \dfrac{-2i\omega_c}{1+k^2/k_{max}^2} & 0 & ck_x & \dfrac{-i\omega_c \Gamma}{1+k^2/k_{max}^2} & \dfrac{-i\omega_p^2}{1+k^2/k_{max}^2} \\ -ck_y & ck_x & 0 & 0 & 0 \\ i & 0 & 0 & -i\Gamma & 0 \\ 0 & i & 0 & 0 & -i\Gamma \end{pmatrix}. \tag{A5}$$

## Appendix B: Quasi-static approximation

In this Appendix, we derive a quasi-static approximation for the dispersion of the edge states at an interface between the Berry dipole material and a metal. It is well-known that quasi-static approximations apply in scenarios where the retardation effects are negligible, allowing the speed of light to be considered effectively infinite ($c \to \infty$). These approximations can be particularly accurate in processes dominated by near-field interactions and evanescent waves.

In the quasi-static limit the electric and magnetic fields are effectively decoupled, thereby simplifying Maxwell's equations. In such a context, the electric field can be written as a gradient of an electric potential $\phi$:

$$\mathbf{E} = -\nabla \phi. \tag{B1}$$

From Gauss' law $\nabla \cdot \mathbf{D} = 0$ and from the constitutive relation $\mathbf{D} = \varepsilon_0 \bar{\varepsilon} \cdot \mathbf{E}$, it follows that the electric potential is constrained as:

$$\nabla \cdot \left(\bar{\varepsilon} \cdot \nabla \phi\right) = 0. \tag{B2}$$

We use the following ansatz for the electric potential,

$$\phi = \phi_0 e^{ik_x x} \begin{cases} e^{-\gamma_{QS} y}, & y > 0 \\ e^{+\gamma_{m,QS} y}, & y < 0 \end{cases}, \tag{B3}$$

to describe a surface wave propagating at the interface ($y = 0$) between the Berry dipole material and a metal (Fig. 1b). Here, $\gamma_{QS}, \gamma_{m,QS}$ represent the attenuation constants along the direction perpendicular to the interface, and $k_x$ is the propagation constant of the surface wave (plasmons). Note that the ansatz ensures that the electric potential is continuous at the interface.

In order to determine $\gamma_{QS}, \gamma_{m,QS}$, we note that from Gauss' law [Eq. (B2)], the dispersion of a plane wave in a bulk region ($\phi \sim e^{ik_x x} e^{ik_y y}$) reduces to:

$$\mathbf{k} \cdot \bar{\varepsilon} \cdot \mathbf{k} = 0. \tag{B4}$$

As the anti-symmetric part of the permittivity does not contribute to the scalar $\mathbf{k} \cdot \bar{\varepsilon} \cdot \mathbf{k}$, it can be readily shown using the above equation that both attenuation constants are identical and independent of the permittivity tensor:

$$\gamma_{QS} = \gamma_{m,QS} = |k_x|. \tag{B5}$$

Enforcing now the continuity of the normal component of the electric displacement vector $\hat{\mathbf{y}} \cdot \mathbf{D}(0^+) = \hat{\mathbf{y}} \cdot \mathbf{D}(0^-)$, we obtain the quasi-static dispersion of the edge modes:

$$-\varepsilon_{EO} k_x - \varepsilon_D |k_x| = \varepsilon_m |k_x|. \tag{B6}$$

Plugging the expressions of $\varepsilon_m, \varepsilon_{EO}, \varepsilon_D$ into the above equation, one obtains the following cubic equation:

$$\omega^3 + \omega^2 (s\omega_c + i\Gamma) + \omega \left( -\frac{\omega_p^2}{2} - \frac{\omega_m^2}{2} + i\frac{3}{2}\Gamma s\omega_c \right) - i\frac{\Gamma}{2}\omega_m^2 = 0, \tag{B7}$$

with $s = \mathrm{sgn}(k_x)$.

–28–

The quasi-static solutions of Eq. (B7) give the asymptotic values of the edge states dispersion when $k_x \to \infty$, specifically, $\omega_{\text{QS,edge}} = \lim_{k_x \to \infty} \omega_{\text{exact}}(k_x)$. The quasi-static solutions are marked in Fig. 4 with black triangular points, and match precisely the solutions of the exact dispersion [Eq. (11)] in the short-wavelength limit. There are a total of six solutions because the parameter *s* can take the values $s = \pm 1$. In the PEC limit ($\omega_m \to \infty$), the only finite quasi-static solution is $\omega = -i\Gamma$.